\pdfoutput=1

\documentclass[11pt]{article}

\usepackage{acl}

\usepackage{times}
\usepackage{latexsym}

\usepackage[T1]{fontenc}

\usepackage[utf8]{inputenc}

\usepackage{color}
\usepackage{tabularx}
\usepackage{tabu}
\usepackage{booktabs}
\usepackage{multirow}
\usepackage{booktabs,tabularx,multirow,xcolor,colortbl,array}
\usepackage{makecell}
\usepackage{threeparttable}
\usepackage{graphicx} 
\usepackage{float}
\usepackage{subfigure} 
\usepackage{amssymb}

\usepackage{hyperref}
\usepackage{tablefootnote}
\usepackage{xspace}

\usepackage{float}
\usepackage{amsmath}
\usepackage{bm}
\usepackage{algorithmicx}
\usepackage{algorithm}
\usepackage{algpseudocode}

\usepackage{arydshln}
\usepackage[dvipsnames]{xcolor}
\usepackage[table]{xcolor}
\definecolor{rowblue}{HTML}{F4F5F8} 

\usepackage{amssymb}
\usepackage{pifont}
\usepackage{enumitem}
\usepackage{forest}
\usepackage{adjustbox}
\usepackage{graphicx}
\usetikzlibrary{arrows.meta,positioning}
\useforestlibrary{edges}

\usepackage{microtype}

\newcommand{\bright}{\textsc{Bright}\xspace}

\newcommand{\eg}{\hbox{\emph{e.g.,}}\xspace}
\newcommand{\ie}{\hbox{\emph{i.e.,}}\xspace}
\usepackage{array} 
\newcolumntype{L}[1]{>{\raggedright\arraybackslash}p{#1}}

\newcommand{\benchname}[2]{\shortstack[l]{#1\\[-0.35ex]{\scriptsize\cite{#2}}}}

\providecommand{\papersep}{;\allowbreak\ }
\providecommand{\paperlist}[1]{%
  \footnotesize
  \parbox[t]{\linewidth}{%
    \raggedright 
    \setlength{\parindent}{0pt}%
    \setlength{\parskip}{0pt}%
    #1%
  }%
}

\title{A Survey of Reasoning-Intensive Retrieval: Progress and Challenges}

\author{
  Yiyang Wei$^1$\thanks{~~Equal contributions. Correspondence to: Tingyu Song (\texttt{songtingyu23@mails.ucas.ac.cn}), Yilun Zhao (\texttt{yilun.zhao@yale.edu}).} \quad 
  Tingyu Song$^{2*}$ \quad
  Siyue Zhang$^{3}$ \quad
  Yilun Zhao$^{4}$  \\[3pt]
  $^{1}$Zhejiang University \quad $^{2}$University of the Chinese Academy of Sciences \\ $^{3}$Nanyang Technological University \quad $^{4}$Yale University \\
}
\begin{document}
\maketitle
\begin{abstract}
Reasoning-Intensive Retrieval (RIR) targets retrieval settings where relevance is mediated by latent inferential links between a query and supporting evidence, rather than semantic similarity. 
Motivated by the emergent reasoning abilities of Large Language Models (LLMs), recent work integrates these capabilities into the IR field, spanning the entire pipeline from benchmarks to retrievers and rerankers. 
Despite this progress, the field lacks a systematic framework to organize current efforts and articulate a clear path forward. 
To provide a clear roadmap for this rapidly growing yet fragmented area, this survey (1) systematizes existing RIR benchmarks by knowledge domains and modalities, providing a detailed analysis of the current landscape; 
(2) introduces a structured taxonomy that categorizes methods based on where and how reasoning is integrated into the retrieval pipeline, alongside an analysis of their trade-offs and practical applications; 
and (3) summarizes challenges and future directions to guide research in this evolving field. 
\end{abstract}

\section{Introduction}
Information Retrieval (IR) underpins everyday information access (\ie web search) and has advanced rapidly in real world applications~\cite{devlin-etal-2019-bert, izacard2022unsupervised}. 
Within the rise of deep research and agentic search~\cite{qiao2025webresearcherunleashingunboundedreasoning, youtu_agent}, 
retrieval has increasingly extended to more scenarios such as multi-hop~\cite{yang-etal-2018-hotpotqa}, instruction-following~\cite{weller-etal-2025-followir, weller2025promptriever}, and long-context retrieval~\cite{zhu-etal-2024-longembed, saad2024benchmarklongcontext}. 

\begin{figure}[t] 
    \centering
    \includegraphics[width=\linewidth]{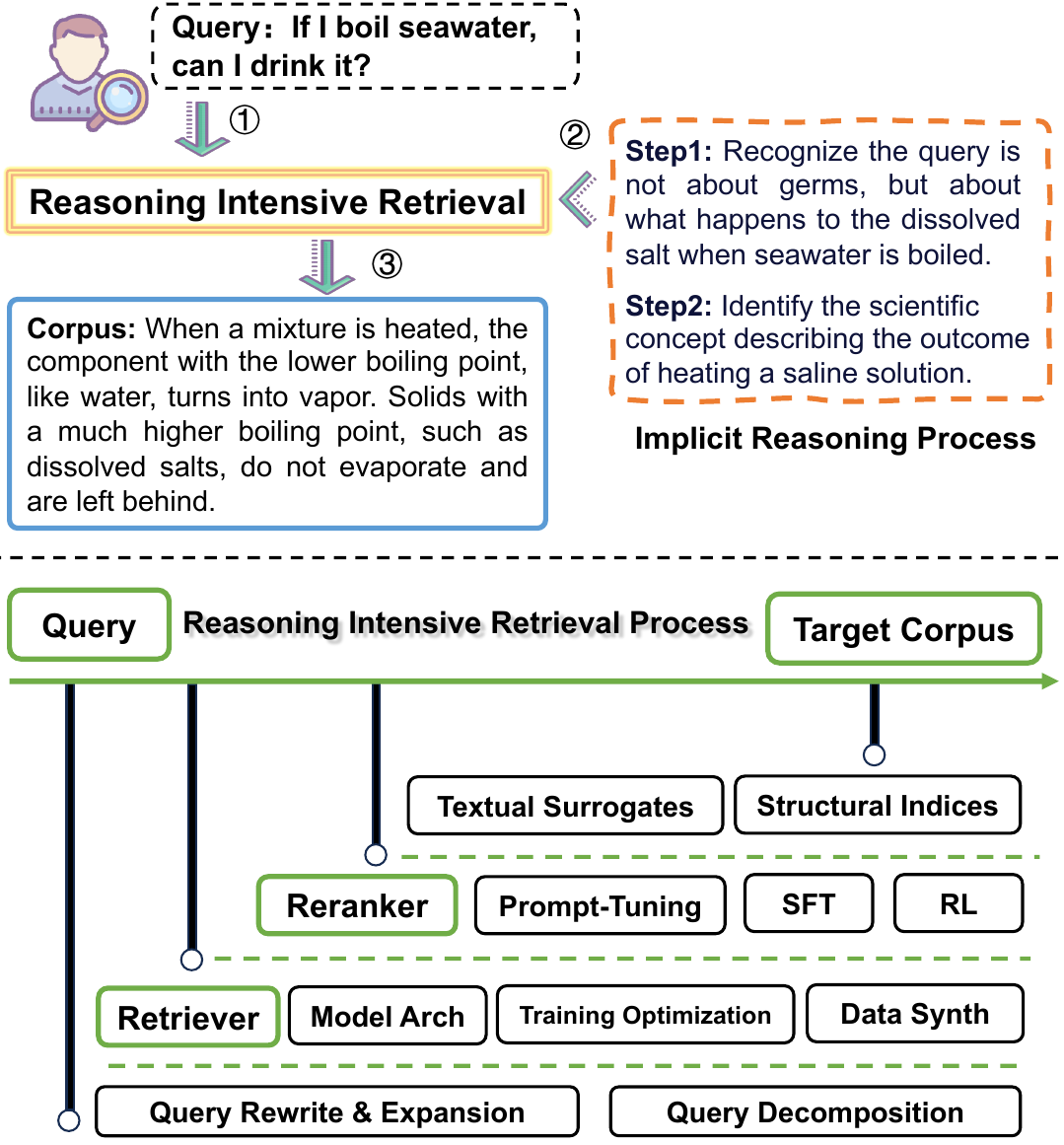}
    \caption{\textbf{Top:} An example of reasoning-intensive retrieval, where a query and its supporting document are connected through an implicit multi-hop reasoning chain. \textbf{Down:} Overview of the retrieval pipeline and representative techniques, which is detailed in Section~\ref{sec:method}.}
    \label{fig:methodologies}
\end{figure}

These advances aim for scenarios with high semantic overlap. 
However, retrieval in expert domains requires not just overcoming lexical or semantic distances, but a deeper reasoning capability to infer implicit connections, such as mapping a brief algorithm description to its symbolic code. We refer to this setting as 
\textbf{\underline{R}}easoning-
\textbf{\underline{I}}ntensive 
\textbf{\underline{R}}etrieval (\textbf{RIR})
where relevance is based on latent inferential links connecting a query to supporting evidence. For example, as shown in \autoref{fig:methodologies}, answering whether boiled seawater is drinkable requires retrieving evidence about the behavior of dissolved salt during boiling, even though the query and the relevant document are linked only through an implicit multi-hop reasoning chain rather than direct lexical overlap.

\providecommand{\paperlist}[1]{\footnotesize\begin{tabular}{@{}l@{}}#1\end{tabular}}
\providecolor{hidden-black}{rgb}{0,0,0}

\begin{figure*}[t!]
    \centering
    \resizebox{0.98\textwidth}{!}{%
        \begin{forest}
            leaf/.style={
                fill=green!10,
                draw,
                text=black,
                align=left,
                font=\normalsize,
                inner xsep=0.25em,
                inner ysep=0.25em,
            },
            forked edges,
            for tree={
                child anchor=west,
                parent anchor=east,
                grow'=east,
                anchor=west,
                base=left,
                font=\small,
                rectangle,
                draw=hidden-black,
                rounded corners,
                minimum height=2em,
                minimum width=4em,
                edge+={darkgray, line width=1pt},
                s sep=3pt,
                inner xsep=0.4em,
                inner ysep=0.6em,
                line width=0.8pt,
                text width=8em,
                align=center,
                l sep=6pt,
                where level=1{text width=6em, font=\small}{},
                where level=2{text width=8em}{},
                where level=3{where n children=0{leaf, text width=46em}{}}{},
                where level=4{where n children=0{tikz={leaf}, text width=36.5em}{}}{},
                where level>=5{where n children=0{tikz={leaf}, text width=30em}{}}{},
                ver/.style={
                    fill=white!50,
                    rotate=90,
                    child anchor=north,
                    parent anchor=south,
                    anchor=center,
                    text width=8em,
                    font=\bfseries,
                },
            },
            [Taxonomy, ver
                [Benchmarks \S\ref{bench}, fill=blue!18
                    [Open Domain \S\ref{bench:opendomain}, fill=blue!10
                        [\paperlist{ImpliRet \cite{taghavi-etal-2025-impliret}\papersep BESPOKE \cite{kim2025bespoke}}, tikz={leaf}]
                    ]
                    [Expert Domain \S\ref{bench:expertdomain}, fill=blue!10
                        [\paperlist{MIRB \cite{ju2025mirb}\papersep MathNet-Retrieve \cite{alshammari2025mathnet}\papersep ScIRGen \cite{lin2025scirgen}\papersep FreshStack \cite{thakur2025freshstack}\papersep CoIR \cite{li-etal-2025-coir}\papersep CoQuIR \cite{geng2025coquir}\papersep LegalBenchmark \cite{legalbenchmark2025}\papersep R2MED \cite{li2025r2med}\papersep CMIRB \cite{li-etal-2025-automir}, etc.}, tikz={leaf}]
                    ]
                    [Multi-Domain \S\ref{bench:multidomain}, fill=blue!10
                        [\paperlist{\bright~\cite{su2025bright}\papersep \bright-Plus \cite{chen2025brightplus}\papersep RAR-b \cite{xiao2024rarb}}, tikz={leaf}]
                    ]
                    [Multimodal \S\ref{bench:multimodaldomain}, fill=blue!10
                        [\paperlist{MRMR \cite{zhang2025mrmr}\papersep MR2-BENCH \cite{zhou2025mr}\papersep ARK \cite{lin2026arkdualaxismultimodalretrieval}\papersep MM-BRIGHT \cite{abdallah2026mmbrightmultitaskmultimodalbenchmark}}, tikz={leaf}]
                    ]
                ]
                [Methods \S\ref{method}, fill=red!18
                    [Pre-retrieval \S\ref{method:preretrieval}, fill=red!10
                        [Query Reform \& \\Decompose \S\ref{methodsec:queryreform}, fill=yellow!10
                            [\paperlist{TongSearch-QR \cite{qin-etal-2025-tongsearch};\ ConvSearch-R1 \cite{zhu-etal-2025-convsearch}\papersep ThinkQE \cite{lei-etal-2025-thinkqe};\ DIVER-QExpand \cite{long2025diver}\papersep RAR2 \cite{xu-etal-2025-rar2};\ RITE \cite{rite2024exploring};\ AdaQR \cite{zhang2025youradaqr};\ ReDI \cite{zhong2025reasoningdecomp}\papersep The Logical Retrieval System \cite{faltings-etal-2025-enhancing}\papersep LaSER~\cite{jin2026laserinternalizingexplicitreasoning}}, tikz={leaf}]
                        ]
                        [Index Enrich \S\ref{methodsec:docindex}, fill=yellow!10
                            [\paperlist{SPIKE \cite{lee2025imaginespike}\papersep EnrichIndex \cite{chen2025enrichindex}\papersep Representation Sharpening for Zero-Shot Dense Retrieval \cite{ashok2025representation}\papersep LATTICE \cite{gupta2025llmlattice}\papersep Reranker-Guided Search \cite{rerankerguided2025}}, tikz={leaf}]
                        ]
                    ]
                    [Retriever Training \S\ref{method:retriever}, fill=red!10
                        [Base Model \S\ref{methodsec:basemodel}, fill=yellow!10
                            [\paperlist{LLM2Vec \cite{behnamghader2024llmvec};\ Gecko \cite{lee2024gecko};\ DLM \cite{zhang-etal-2025-diffusion}}, tikz={leaf}]
                        ]
                        [Training Data \S\ref{mehodsec:trainingdata}, fill=yellow!10
                            [\paperlist{ReasonIR \cite{shao2025reasonir};\ RaDeR \cite{das-etal-2025-rader};\ ReasonEmbed \cite{chen2025reasonembed}\papersep SQUARE \cite{yoon-etal-2025-square};\ DIVER \cite{long2025diver}}, tikz={leaf}]
                        ]
                        [Objective \& \\Reward \S\ref{methodsec:objectives}, fill=yellow!10
                            [\paperlist{LREM \cite{tang2025largelrem};\ O1-Embedder \cite{yan2025o1embed};\ UME-R1 \cite{lan2025ume}\papersep Revela \cite{cai2025reveladenseretrieverlearning}}, tikz={leaf}]
                        ]
                    ]
                    [Reranking \S\ref{method:rerank}, fill=red!10
                        [Prompt-Tuning \S\ref{methodsec:prompttuning}, fill=yellow!10
                            [\paperlist{InsertRank \cite{seetharaman2025insertrank}\papersep JudgeRank \cite{niu2024judgerank}}, tikz={leaf}]
                        ]
                        [SFT \& Distill \S\ref{methodsec:sftdistill}, fill=yellow!10
                            [\paperlist{LimRank \cite{song-etal-2025-limrank};\ ERank \cite{cai2025erank}\papersep ReasonRank \cite{liu2025reasonrank};\ GroupRank \cite{sun2025grouprank}\papersep InteRank \cite{samarinas2025interank};\ Reason-to-Rank \cite{ji2025reasontorank}\papersep Rank1 \cite{weller2025rank1};\ Rank-K \cite{yang2025rankk};\ DEAR \cite{abdallah-etal-2025-dear}}, tikz={leaf}]
                        ]
                        [RL \S\ref{methodsec:rl}, fill=yellow!10
                            [\paperlist{Rank-R1 \cite{zhuang2025rank};\ TFRank \cite{fan2025tfrank}\papersep InteRank \cite{samarinas2025interank}\papersep REARANK \cite{zhang-etal-2025-rearank}\papersep ERank \cite{cai2025erank}\papersep GroupRank \cite{sun2025grouprank}\papersep ReasonRank \cite{liu2025reasonrank}}, tikz={leaf}]
                        ]
                    ]
                    [Iterative \S\ref{method:iterativeretrieval}, fill=red!10
                        [\paperlist{Refine Thought \cite{wang2025refine};\ Think Before You Retrieve \cite{vijay2025thinkbefore}\papersep SMR \cite{lee-etal-2025-token};\ CG-Planning \cite{li-etal-2025-elicitcentric}}, tikz={leaf}]
                    ]
                ]
            ]
        \end{forest}%
    }
    \caption{Taxonomy of Reasoning-Intensive Retrieval (RIR).}
    \label{fig:rir-taxonomy}
\end{figure*}
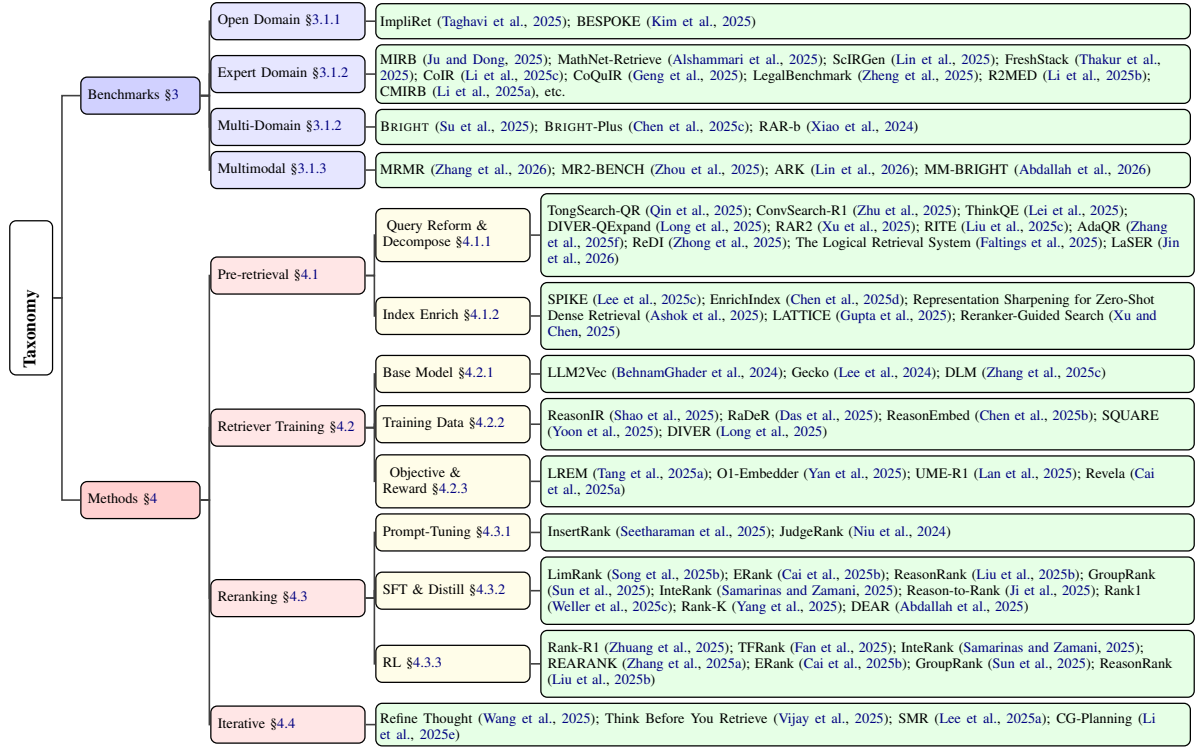

To evaluate the corresponding abilities of current retrieval systems, \bright is introduced as an early benchmark~\cite{su2025bright}. Subsequent efforts have extended RIR evaluation to domain-specific scenarios~\cite{legalbenchmark2025, li2025r2med, ju2025mirb} and  multimodal settings~\cite{zhang2025mrmr, zhou2025mr}, exposing the limitations of state-of-the-art retrievers. 
Motivated by these findings, a growing family of methods integrate reasoning into different stages of the retrieval pipeline, through query-side transformation~\cite{qin-etal-2025-tongsearch, lei-etal-2025-thinkqe, xu-etal-2025-rar2}, reasoning-aware representation learning~\cite{shao2025reasonir, long2025diver, lan2025ume}, and reranking~\cite{song-etal-2025-limrank, zhuang2025rank, liu2025reasonrank}, to improve retrieval performance on reasoning-intensive queries.
Recent studies further suggest that effective RIR may require iterative retrieval pipelines that repeatedly alternate between retrieval and reasoning~\cite{wang2025refine, vijay2025thinkbefore}. 

Despite this rapid progress, existing RIR research still faces two main limitations. First, the evaluation landscape remains highly heterogeneous. Current studies adopt diverse problem formulations, datasets, and evaluation setups across tasks and domains (\eg code, biomedical, math). Second, methodological developments are scattered across different stages of the retrieval pipeline, including query rewriting, retriever training, reranking, and iterative retrieval frameworks. As a result, the field remains difficult to navigate and lacks consistent evaluation and methodological organization. 
In this survey, we aim to address these issues by (1) systematizing existing benchmarks according to reasoning type, domain, and source of difficulty (Section~\ref{bench}); (2) proposing a structured taxonomy of RIR methods based on where reasoning is introduced in the retrieval pipeline (Section~\ref{method}), and analyzing their trade-offs and application scenarios (Appendix~\ref{app:method_comparison}); and (3) outlining key open challenges in evaluation metrics, domain generalization, inference cost, and multimodal reasoning (Section~\ref{challengesandfuture}). 

\section{Related Work}
Reasoning-intensive Retrieval (RIR) is a nascent but rapidly emerging domain. However, to the best of our knowledge, comprehensive surveys of this field are still scarce. 
Existing IR surveys have made substantial contributions in cataloguing the evolution of retrieval paradigms~\cite{bm25surveyrobertson,yates-etal-2021-pretrained, generativesurveyli, Zhang2025OnTR}, but these works primarily focus on semantic or lexical query-document relevance, leaving the inferential demands placed on the retrieval system largely unaddressed. 
When it comes to the intersection between reasoning and retrieval,  current surveys often emphasize the role of reasoning within RAG and agentic frameworks, such as RAG-Reasoning~\cite{li-etal-2025-survey-rag} and Reasoning Agentic RAG~\cite{liang2025reasoningrag12}, they typically treat retrieval as a preliminary stage to support generation. These works prioritize how to leverage retrieved evidence for reliable answers rather than the inferential depth of the retrieval process itself. In contrast, RIR focuses on the retrieval system's intrinsic ability to infer connections between a query and the target corpus through implicit logical inferential links~\cite{su2025bright, zhang2025mrmr}. In this setting, retrieval is the end task under a framework of inference-mediated relevance.

\section{Reasoning-Intensive IR Evaluation}
\label{bench}
In this section, we compile existing benchmarks for reasoning-intensive retrieval and provide a comparative analysis across them.
\begin{table}[t]
\centering
\resizebox{\linewidth}{!}{
\small
\begin{tabular}{@{}l l@{\hspace{2pt}}r l@{}}
\toprule
\textbf{Domain} & \textbf{Name} & \textbf{Size} & \textbf{Annotation Type} \\
\midrule
Open Domain  & ImpliRet        & 9,000   & LLM-Automated \\
             & BESPOKE         & 150     & Human-Curated \\
\midrule
Scientific   & MIRB            & 39,029  & Derived$^1$ \\
             & MathNet-Retrieve         & 10,000  & Hybrid$^2$ \\
             & SciRGen         & 61,376  & LLM-Automated \\
             & FreshStack      & 672     & LLM-Automated \\
\midrule
Code         & CoIR            & $\approx$162,000 & Derived \\
             & CoQuiR          & 42,725  & LLM-Automated \\
\midrule
Legal        & Legal-Benchmark & 9,863   & Human-Curated \\
\midrule
Medical      & R2MED           & 876     & Hybrid \\
             & CMIRB         & 10,962  & LLM-Automated \\
\midrule
Multi-Domain & \bright          & 1,384   & Hybrid \\
             & \bright-Plus         & 1,384   & Hybrid \\
             & RAR-b           & 45,745  & Derived \\
\midrule
Multi-Modal  & MRMR            & 1,435   & Hybrid \\
             & MR2-BENCH       & 1,309   & Hybrid \\
             & ARK             & 1,547   & Hybrid \\
             & MM-BRIGHT       & 2,803  & Hybrid  \\
             
\bottomrule
\end{tabular}
}
\caption{Summary of RIR evaluation Benchmarks (see full table in \autoref{tab:benchmarks} in Appendix). 
$^1$Derived: Source is derived from established data sources (\eg previous datasets, libraries, internet QA). $^2$Hybrid: Source is both LLM-Automated and Human-Curated.  
}
\label{tab:tight_benchmark}
\end{table}
\subsection{Existing Evaluation Benchmarks}
Current reasoning-intensive retrieval benchmarks cover a broad range of domains. We classify them into the following four types:
(1) \textit{open-domain}, which covers general-purpose knowledge and commonsense reasoning;
(2) \textit{expert-domain}, which probes specialized knowledge within a single professional discipline;
(3) \textit{multi-domain}, which aggregates tasks from multiple professional areas to test knowledge breadth;
(4) \textit{multimodal}, which introduces unique challenges distinct from text-only processing and represents a significant frontier. We first provide a brief summary of these benchmarks in \autoref{tab:tight_benchmark}, and present a more comprehensive overview in \autoref{tab:benchmarks} in the Appendix. We next detail these evaluation benchmarks:

\subsubsection{Open-Domain Benchmarks}
\label{bench:opendomain}
Open-domain benchmarks operate on general-purpose knowledge and commonsense, without requiring specialized expertise. The primary reasoning challenge in these daily settings is to decipher the user's latent intent, which is often implicit and context-dependent. 
To this end, the BESPOKE~\cite{kim2025bespoke} and ImpliRet~\cite{taghavi-etal-2025-impliret} benchmarks construct evaluation frameworks using user chat histories, where queries are frequently short and ambiguous. 
They pose a significant challenge to current models by explicitly testing their ability to recover underlying intent from the conversational context, providing a realistic measure of current models' practical utility. 

\subsubsection{Expert-Domain Benchmarks} 
\label{bench:expertdomain}
Expert-domain benchmarks address professional fields where specialized knowledge and domain-specific practices complicate relevance assessment, necessitating reasoning abilities beyond what is required in general settings. 

\paragraph{Scientific.}
The scientific domain encompasses fields built on formal systems of knowledge. 
For instance, ScIRGen~\cite{lin2025scirgen} addresses the lack of realism in scientific QA benchmarks by proposing a scalable generation framework that creates complex, task-implicit questions grounded in papers. 
FreshStack~\cite{thakur2025freshstack} is the first to deliver an automated retrieval evaluation benchmark tailored to real developer needs in technical documentation domain. 
In mathematics, MIRB~\cite{ju2025mirb} and MathNet-Retrieve~\cite{alshammari2025mathnet} evaluate whether systems can retrieve mathematically relevant statements. While MathNet-Retrieve~\cite{alshammari2025mathnet} focuses on equivalent problems across multilingual and multimodal contexts, MIRB\cite{ju2025mirb} extends the evaluation to more reasoning tasks, including theorem-level premise retrieval and problem-solving answer retrieval.

\paragraph{Legal.}
Legal retrieval is challenging because it requires bridging abstract legal rules with concrete, case-specific situations. 
This challenge extends to precedent retrieval, which involves identifying legally analogous cases that share overlapping legal principles~\cite{nigam2022coliee, li2023muser}. 
To evaluate this reasoning capability directly, a new benchmark~\cite{legalbenchmark2025} introduces two reasoning-intensive tasks, Bar Exam QA and Housing Statute QA, which require systems to connect factual scenarios to their governing statutes through analytical and deductive reasoning. 

\paragraph{Medical.}
In medicine, a similar challenge arises, but the ambiguity stems not from abstract rules but from underspecified, symptom-centered queries. 
Benchmarks like R2MED~\cite{li2025r2med} and CMIRB~\cite{li-etal-2025-automir} evaluate retrieval for vague patient presentations, where relevance is determined by linking symptoms to plausible diagnoses and appropriate treatment plans. 

\paragraph{Code. }
Compared with natural-language retrieval, reasoning-intensive retrieval in code demands reasoning over symbols and structure. 
CoIR~\cite{li-etal-2025-coir}, for instance, assesses a model's ability to reason about program behavior through tasks like cross-language code equivalence and bug localization. 
Building on this, CoQuIR~\cite{geng2025coquir} pushes further by demanding that retrievers discriminate not only by functionality but also by code quality, with attributes including correctness, efficiency, and security. 
These benchmarks signal a shift from retrieving topically relevant code~\cite{husain2020codesearchnetchallengeevaluatingstate} to identifying high-quality, reliable solutions.

\paragraph{Multi-Domain Benchmarks. }
\label{bench:multidomain}
In contrast to benchmarks focused on a single domain, multi-domain benchmarks aggregate representative tasks from several professional fields to provide a broader evaluation of current models' capabilities.  
For example, \bright~\cite{su2025bright} and \bright-Plus~\cite{chen2025brightplus} exemplify this direction by covering specialized areas such as science, technology, engineering, and mathematics, and by including queries on topics such as software debugging and scientific theorem retrieval. 
Meanwhile, RAR-b~\cite{xiao2024rarb} derives retrieval instances from multiple-choice QA to probe diverse reasoning skills (\eg commonsense, temporal), but its shorter retrieval targets make it closer to conceptual capability testing than document-level professional search. 

\subsubsection{Multimodal Benchmarks}
\label{bench:multimodaldomain}
Multimodal RIR benchmarks introduce novel challenges by moving beyond text-only retrieval to tasks that demand reasoning across diverse modalities (\eg image, text). 
Recent multimodal retrieval benchmarks, including MRMR~\cite{zhang2025mrmr}, MM-BRIGHT~\cite{abdallah2026mmbrightmultitaskmultimodalbenchmark}, and ARK~\cite{lin2026arkdualaxismultimodalretrieval}, introduce reasoning-heavy and knowledge-intensive tasks that require models to capture abstract conceptual connections across scientific multimodal documents and diverse domains. In contrast,  MR$^2$-Bench~\cite{zhou2025mr} broadens the task scope and places stronger emphasis on evaluating spatial, logical, and causal reasoning capabilities through challenging scenarios such as visual puzzles and dimensional transformations.

\subsection{Comparative Benchmark Analysis}
Having reviewed the landscape of reasoning-intensive IR benchmarks across domains and modalities, we now turn to a comparative analysis that highlights two key axes: the scale-reliability trade-off in benchmark construction and emphasis on different reasoning types across domains.
\paragraph{Scale–Reliability Trade-offs. } 
A fundamental trade-off exists in benchmark design between scalable synthetic generation and rigorous human curation (see \autoref{tab:tight_benchmark} and \autoref{tab:benchmarks}). 
On one hand, LLM-based synthetic benchmarks like ScIRGen~\cite{lin2025scirgen} and ImpliRet~\cite{taghavi-etal-2025-impliret} expand coverage and diversify cognitive demands, but can suffer from hallucinations and limited validation.
On the other hand, reliability-oriented benchmarks emphasize human/expert oversight, including \bright~\cite{su2025bright} and its cleaned extension \bright-Plus~\cite{chen2025brightplus} sources data from human experts across various domains to ensure trustworthiness. This emphasis on reliability becomes paramount in high-stakes fields, such as Bar Exam QA~\cite{legalbenchmark2025}.
Thus, an open direction is hybrid construction pipelines that scale via synthesis while preserving evaluative validity through targeted expert checks. 

\paragraph{Reasoning Types and Domain Emphases.}
\begin{table}[t]
\centering
\small
\label{tab:reasoning-type-definitions}
\begin{tabular}{p{0.93\linewidth}}
\toprule
\textbf{Deductive Reasoning}: A general principle or theorem in the document is directly applied to explain a specific scenario or solve a problem in the query. \\\midrule

\textbf{Analogical Reasoning}: A document draws a parallel with the query in its underlying logic, indicating that the query and document share a solution strategy or a common theorem/algorithmic foundation. \\\midrule

\textbf{Causal Reasoning}: The document identifies root causes or mechanistic relationships that explain effects observed in the query. Resolution requires tracing causal chains from symptoms to origins. \\\midrule

\textbf{Analytical Reasoning}: The document provides critical domain knowledge that fills gaps in multi-step reasoning chains required to resolve the query. This involves decomposition of complex problems into interdependent sub-questions. \\\midrule

\textbf{Numerical Reasoning}: The query is resolved by applying quantitative constraints in the document, requiring arithmetic computation (\eg percentages, unit conversion, rate/ratio) or time arithmetic (\eg duration, scheduling offsets, temporal comparisons). The logical mechanism is a deterministic mapping from numeric facts and rules to a target value or decision. \\
\bottomrule
\end{tabular}
\caption{Definitions of the five reasoning types covered by existing RIR benchmarks.}
\label{tab:reasoning-type}
\end{table}
Following \bright~\cite{su2025bright}, we categorize RIR benchmarks into five reasoning types---\emph{deductive}, \emph{analogical}, \emph{causal}, \emph{analytical}, and \emph{numerical}, as summarized in \autoref{tab:reasoning-type}. We provide representative examples and inference chains for each type in \autoref{tab:domain-benchmarks} in Appendix.
\emph{Numerical reasoning} often involves arithmetic or temporal operations in daily settings~\cite{taghavi-etal-2025-impliret}, whereas \emph{deductive reasoning} is the most prevalent across domains, supporting rule-to-case application in mathematics/science~\cite{ju2025mirb}, medicine~\cite{li2025r2med,li-etal-2025-automir}, and law~\cite{legalbenchmark2025}.
\emph{Analogical reasoning} is particularly salient in code~\cite{li-etal-2025-coir,su2025bright} and math~\cite{alshammari2025mathnet, ju2025mirb} benchmarks for establishing functional correspondences across modalities.
Finally, \emph{causal} and \emph{analytical} reasoning frequently appear in specialized tasks such as troubleshooting and problem decomposition.

\section{Reasoning-Intensive IR Methods}\label{sec:method}
\label{method}
Reasoning-intensive retrieval can inject reasoning at different points of the retrieval pipeline, from shaping the input to refining candidates during ranking and multi-step interaction.
To make these design choices comparable, we organize existing methods by \emph{where} reasoning is introduced and \emph{how} it interacts with retrieval.
Accordingly, we structure this section into four stages: pre-retrieval augmentation, retrieval, reranking, and iterative workflows.
To complement this structural taxonomy, Appendix~\ref{app:method_comparison} provides a comparative analysis across these categories, while \autoref{app:application_task} maps the methods to specific downstream tasks and applications.
\subsection{Pre-Retrieval Reasoning Augmentation}
\label{method:preretrieval} 
To enhance RIR, pre-processing techniques can be applied to both queries and documents before the matching stage. Query-side methods (\S\ref{methodsec:queryreform}) focus on refining or decomposing user's request to clarify its underlying intent; document-side methods (\S\ref{methodsec:docindex}) aim to enrich the document corpus, making latent evidence more explicit and accessible. 
\subsubsection{Query-Side Augmentation}
\label{methodsec:queryreform}
Query-side augmentation methods can be broadly grouped into the following two categories:
\paragraph{Query Rewriting and Expansion.} Query rewriting and expansion leverages LLM-generated reasoning traces to reformulate or enrich the original query, aiming to make the underlying information need more explicit for downstream retrieval.
TongSearch-QR~\cite{qin-etal-2025-tongsearch} and ConvSearch-R1~\cite{zhu-etal-2025-convsearch} leverage Reinforcement Learning (RL) with thinking format reward and performance reward to train LLM on query rewriting tasks, achieving better performance with smaller model size.
In addition, ConvSearch-R1~\cite{zhu-etal-2025-convsearch} adopts a cold-start supervised fine-tuning (SFT) stage before RL to improve output format adherence and stabilize reasoning and rewriting behaviors.
For query expansion, RAR2~\cite{xu-etal-2025-rar2} fine-tunes LLMs with a thought dataset and Direct Preference Optimization (DPO) to generate reasoning traces that augment retrieval in clinical scenarios. 
Moving beyond a single-pass expander, ThinkQE~\cite{lei-etal-2025-thinkqe} formulates query expansion as an interactive process that iteratively refines expansions using retrieval feedback, and DIVER-QExpand~\cite{long2025diver} simplifies this workflow by retaining the original and the final-round expansion to control token growth while preserving key information. 
Beyond text-only rewriting, AdaQR~\cite{zhang2025youradaqr} and LaSER~\cite{jin2026laserinternalizingexplicitreasoning} produce latent reasoning in the embedding space, increasing retrieval performance while maintaining low inference latency. Beyond single-vector retrieval, AMER~\cite{chen2025singleembeddingscapturingdiverse} autoregressively generates multiple query embeddings for retrieval, outperforming single-embedding baselines. 
\paragraph{Query Decomposition.} Query decomposition breaks a complex query into sub-queries to better capture multifaceted intents.  
This strategy is particularly relevant to \emph{analytical reasoning} retrieval, where solving the task typically requires a multi-step reasoning chain, that each step can be operationalized as a sub-query. 
ReDI~\cite{zhong2025reasoningdecomp} exemplifies this approach with a three-stage pipeline that performs intent recognition, enriches sub-queries for efficient parallel retrieval, and fuses the retrieved results, leveraging LLM reasoning throughout. 
In contrast, the logical retrieval system~\cite{faltings-etal-2025-enhancing} decomposes natural-language queries into sub-queries connected by logical operators (\eg OR, AND, NOT) and aggregates cosine-similarity signals to better handle compositional constraints. 

\subsubsection{Index-Side Augmentation}
\label{methodsec:docindex}
Complementing query rewriting, index-side augmentation shifts the reasoning burden to offline ingestion by pre-enriching document representations with synthetic metadata. We group existing index-side techniques into the following two types: 

\paragraph{Textual Surrogates.}
Textual-surrogate methods expand each document with auxiliary descriptions that anticipate how users might seek it, while remaining compatible with standard dense retrieval pipelines.
SPIKE~\cite{lee2025imaginespike} instantiates this idea by generating hypothetical retrieval scenarios for each document. Similarly, representation sharpening~\cite{ashok2025representation} strengthens index representations via \emph{document-conditioned} contrastive queries that emphasize distinguishing aspects of a document. 
These methods expose the implicit information needs a document could satisfy, enhancing semantic coverage to better support reasoning-driven inferential links.
Beyond effectiveness, EnrichIndex~\cite{chen2025enrichindex} highlights a practical benefit of such enrichment: by shifting semantic expansion offline, enriched indices can reduce repeated online LLM computation during retrieval, lowering latency and cost.

\paragraph{Structural Indices.}
While textual surrogates improve final performance through additional views, structural indices externalize reasoning pathways by organizing knowledge into interpretable frameworks that retrieval can traverse. 
LATTICE~\cite{gupta2025llmlattice} exemplifies this direction by constructing LLM-guided lattice structures that enable multi-level navigation, capturing implicit dependencies and supporting complex reasoning queries through coarse-to-fine exploration.
Similarly, reranker-guided search~\cite{rerankerguided2025} couples retrieval with downstream ranking signals to steer exploration toward higher-utility regions of the corpus, effectively using structured search trajectories to refine retrieval decisions.

\subsection{Reasoning-Aware Retriever Training}
\label{method:retriever}
To improve the retrievers'  reasoning performance in RIR domain, 
current efforts mainly focus on three aspects: 
(1) \emph{model architecture selction}, current methods implement their algorithm to different architecture of models; 
(2) \emph{data curation}, some works carefully curate training data specialized for RIR;
and (3) \emph{training objectives and reward design} used during optimization.

\subsubsection{Base Model Architecture Selection}
\label{methodsec:basemodel}
Different embedding backbone is a design choice for RIR. 
Motivated by the strong reasoning abilities of LLMs, several works~\cite{behnamghader2024llmvec, lee2024gecko} adapt decoder-style architectures for dense embedding models, yielding LLM-based retrievers. 
However, their unidirectional attention limits the effectiveness of incorporating bidirectional context. 
In contrast, Diffusion Language Model (DLM) embeddings~\cite{zhang-etal-2025-diffusion} leverage bidirectional attention to better integrate surrounding information, which improves reasoning efficiency and embedding performance.

\subsubsection{Training Data Curation}
\label{mehodsec:trainingdata}
Despite the importance of the backbone, training data largely determines which reasoning patterns the model can actually learn to represent. 
Curating specialized training data infused with reasoning elements is an important strategy for boosting the performance of retrievers on logic-heavy queries. 

To curate high quality documents for RIR, 
the central reasoning challenge is supervision mining, positive documents should provide evidence that genuinely supports answering the query~\cite{yoon-etal-2025-square}, while negatives should remain lexically or semantically similar to the query yet be unhelpful to resolve it~\cite{shao2025reasonir, long2025diver}. 
For positives, 
SQUARE~\cite{yoon-etal-2025-square} uses LLM-generated hypothetical answers to retrieve and verify supportive positives.
And to curate hard negatives for RIR, 
ReasonIR~\cite{shao2025reasonir} and DIVER~\cite{long2025diver} perform iterative mining guided by LLM-generated rationales, ReasonEmbed~\cite{chen2025reasonembed} further filters candidates using embedding models with LLM relevance annotations, and RaDeR~\cite{das-etal-2025-rader} leverages MCTS with an LLM to synthesize diverse hard-negative training signals. 

In contrast, for (query, thought, document) triplets, where reasoning is realized through generating retrieval ``thoughts'', the central challenge is to synthesize and retain only those thoughts that provide genuine retrieval utility. 
For instance, O1 Embedder~\cite{yan2025o1embed} addresses this by prompting an expert LLM to produce candidate thoughts and filtering them via a retrieval committee. 
On top of that, LREM~\cite{tang2025largelrem} curates training signals by comparing retrieval outcomes with and without the thought, discarding queries that yield no improvement.

\subsubsection{Training Objectives and Reward Design}

\label{methodsec:objectives}

With reasoning-capable backbones and reasoning-intensive supervision in place, an important step is to choose objectives and rewards that internalize these signals into the retriever’s embedding and ranking behaviors. 
A representative direction is multi-task optimization that jointly trains (1) reasoning generation and (2) embedding discrimination (details of loss functions and analysis are in \autoref{app:loss_function}). 
For example, LREM and O1 Embedder~\cite{tang2025largelrem,yan2025o1embed} combine next-token prediction over intermediate thoughts with contrastive losses, typically via a weighted sum, so that the model learns to ``think'' while remaining a competitive embedder. 
In contrast, the Dense Reasoner~\cite{zhang2025youradaqr} distills the effect of LLM reasoning directly into the embedding space by learning an embedding transformation with an MSE objective that matches LLM-reasoned embeddings. 
Extending joint objectives to multimodal retrieval, UME-R1~\cite{lan2025ume} integrates discriminative contrastive learning with generative objectives defined over reasoning trajectories, together with next-token prediction during cold-start SFT, to support both discriminative and reasoning-driven generative embeddings across modalities. Revela~\cite{cai2025reveladenseretrieverlearning} optimizes the retriever directly via a language-modeling objective with in-batch attention, enabling self-supervised retriever learning without query–document pairs.

\label{methodsec:preferencealign}
Building on the above strategies to enhance the RIR performance, 
RL-based alignment further makes the reasoning trajectory itself an explicit optimization target by shaping it with structured rewards. 
In LREM \cite{tang2025largelrem}, a RL stage scores sampled CoTs with a weighted combination of \emph{generation-side} rewards (\eg format compliance and length control) to encourage structured yet concise trajectories, together with an \emph{embedding-side} retrieval-accuracy reward that favors trajectories producing embeddings with stronger discriminative separation.
Similarly, with both generation-side and embedding-side rewards, UME-R1~\cite{lan2025ume} grounds multimodal representation learning in reasoning trajectories, thereby steering training toward higher-quality reasoning-conditioned multimodal embeddings.

\subsection{Reasoning-Enhanced Reranking} 
\label{method:rerank}
Given the retrieved documents,   
a reranker needs to refine documents' order by evaluating documents from multiple perspectives, which involves deeper reasoning abilities to surface the most useful evidence for the query. 
To clarify how rerankers acquire and strengthen such reasoning ability, we group existing approaches into three paradigms: 
(1) Prompt-Tuning, which is conducted during the inference time, 
(2) Supervised reasoning transfer, which is often realized by SFT and Distillation, 
and (3) Reinforcement Learning, which further improves the general abilities in RIR.

\subsubsection{Prompt-Tuning}
\label{methodsec:prompttuning}
Prompted rerankers elicit reasoning at inference time without parameter updates, making them attractive for rapid deployment and out-of-domain transfer. InsertRank~\cite{seetharaman2025insertrank} inserts BM25 score into the prompt to help reranker reasoning on relevance.  
In addition, JudgeRank~\cite{niu2024judgerank} leverages agentic prompting further decomposing reranking into stages such as query analysis and document analysis, which improves robustness on corresponding queries. 

\subsubsection{Supervised Reasoning Transfer}
\label{methodsec:sftdistill}
While prompted rerankers largely rely on inference-time, supervised transfer aims to \emph{internalize} reasoning behaviors through training on curated supervision. In practice, there are mainly two methods: (1) Supervised Fine-Tuning (SFT), which teaches the reranker to make ranking decisions across retrieved passages (\eg relevance scores, orderings), and (2) Reasoning Distillation, which is achieved by training the student to mimic the structured intermediate rationales that the teacher model generates to justify its ranking decisions.

\paragraph{Supervised Fine-Tuning. }
From a pointwise perspective, LimRank~\cite{song-etal-2025-limrank}  generates positive/negative documents derived from long CoT answers to capture implicit relationships between documents and queries.  
However, ERank~\cite{cai2025erank} argues that binary relevance training leads to poor score discrimination, and replaces it with generative SFT that outputs fine-grained integer scores to better separate subtly different candidates. 
Beyond traditional SFT, a cold-start SFT teaches reranker an output format, for instance, reasoning patterns (\texttt{<think>} and \texttt{<answer>})~\cite{liu2025reasonrank} and within-group comparison format~\cite{sun2025grouprank}.

\paragraph{Distillation. }
InteRank~\cite{samarinas2025interank} and Reason-to-Rank~\cite{ji2025reasontorank} improve reasoning skills by distilling ranking explanations from a teacher-LLM, emphasizing that generating explanations is crucial for effective ranking. 
Rank1~\cite{weller2025rank1} and Rank-K~\cite{yang2025rankk} distill reasoning traces into smaller rerankers and enable longer inference-time CoT for the reasoning intensive retrieval queries, yielding stronger performance on \bright. 
At a finer granularity of reasoning, DeAR~\cite{abdallah-etal-2025-dear} introduces token-level relevance distillation, achieving high accuracy on rerank task.

\subsubsection{Reinforcement Learning }
\label{methodsec:rl}
Building on SFT and distillation that largely imitate labeled preferences, RL further aligns both \emph{what} the model ranks and \emph{how} it justifies decisions by optimizing task-level rewards tied to ranking quality, output structure, and explanation usefulness. Recent rerankers largely share a Group Relative Policy Optimization (GRPO) backbone, but they diverge in \emph{how} the reward specifies the target behavior, ranging from strict rule checks to richer, metric-driven objectives.
At the minimalist end, Rank-R1~\cite{zhuang2025rank} uses a strict rule-check reward on the best-document label, enabling reranker reasoning with only a small amount of reasoning-free labeled data.
In contrast, InteRank~\cite{samarinas2025interank} automatically generates reward value from a reasoning LLM-based reward model.
Beyond single-objective rewards, composite rewards that jointly enforce  optimizing ranking from multiple aspects.
For instance, REARANK~\cite{zhang-etal-2025-rearank} and TFRank~\cite{fan2025tfrank} combine a score-based reward and a format reward to encourage better-structured, reasoning-centric outputs.
To inject broader ranking awareness, ERank~\cite{cai2025erank} and GroupRank~\cite{sun2025grouprank} augment pointwise scoring with listwise-derived rewards computed over the entire candidate list (or groups), encouraging the scorer to respect global ordering.
Finally, moving beyond one-shot ranking, ReasonRank~\cite{liu2025reasonrank} optimizes a \emph{multi-view} ranking reward that accounts for the multi-turn nature of sliding-window listwise ranking (combining signals and ranking-similarity measures), so RL explicitly refines end-to-end list quality rather than single-window gains.

\subsection{Reasoning-Driven Iterative Retrieval}
\label{method:iterativeretrieval}

Reasoning injected into a single retrieval stage can improve performance on reasoning-intensive IR tasks, but naively chaining multiple reasoning modules may amplify redundant ``overthinking'' and introduce misaligned or drifting reasoning traces. Consequently, \emph{reasoning-driven iterative retrieval} has emerged as a way to coordinate reasoning across stages, refining the search process through adaptive iterations.
SMR~\cite{lee-etal-2025-token}, for example, enforces a state-machine structure that moves from granular token-level analysis to explicit retrieval actions (\eg \textsc{Refine}, \textsc{Rerank}, \textsc{Stop}). 
Similarly, both ~\citet{li-etal-2025-elicitcentric} and ~\citet{vijay2025thinkbefore} cast retrieval as a test-time, iterative decision process guided by an LLM; notably, ~\citet{vijay2025thinkbefore} implements this guidance as an RL-trained multi-turn retrieval policy with turn-level rewards and reports stronger effectiveness even with a smaller LLM backbone.
In an end-to-end setting, \citet{wang2025refine} propose approaches where embedding models iteratively infer and retrieve within the model, progressively sharpening relevance for complex queries, without necessarily retraining the retriever for each refinement step.

\section{Open Challenges and Future Directions}
\label{challengesandfuture}
Despite rapid progress in RIR, this section examines remaining challenges and future directions.
\paragraph{Evaluation Overly Relies on Traditional IR Metrics.}
Current evaluation protocols~\cite{su2025bright, li2025r2med} still rely primarily on conventional IR metrics such as nDCG and Recall. This introduces two limitations: (1) Efficiency is largely overlooked. Some methods~\cite{long2025diver, chen2025reasonembed} achieve strong effectiveness through complex frameworks but incur high computational costs. 
Recently, some studies~\cite{er2metric} have proposed evaluating both the efficiency and effectiveness of current rerankers, while others~\cite{Zhou2024BeyondCR, weller-etal-2025-followir, song-etal-2025-ifir} have introduced metrics tailored to instruction-following retrieval. Moving forward, however, we will likely need novel metrics specifically designed for reasoning-intensive scenarios (\eg DeepResearch). 
(2) Fine-grained relevance is not well captured. Two models may obtain similar nDCG scores while retrieving qualitatively different results. Thus, metrics that jointly consider effectiveness and efficiency, as well as fine-grained relevance assessment, are promising  directions~\cite{zhang2025qwen3embeddingadvancingtext}.

\paragraph{The Domain Generalization Gap in Evaluation.}
Most RIR benchmarks are anchored in specialized professional settings, from STEM~\cite{su2025bright}, to legal~\cite{legalbenchmark2025} and medical benchmarks~\cite{li-etal-2025-automir}. 
Although these resources provide evidence-rich scenarios for structured reasoning (see \autoref{tab:domain-benchmarks}), their distance from everyday information needs limits their coverage of broader retrieval tasks.
Recent works~\cite{kim2025bespoke, taghavi-etal-2025-impliret} take a step in this direction by testing intent resolution from implicit queries over chat histories, but limited in scale and task diversity.
A key next step is scalable, heterogeneous evaluation with broader coverage and stronger generalizability, grounded in routine human--AI interactions.

\paragraph{Bridging the Multimodal Reasoning Gap.}
Most existing RIR research is confined to text-only~\cite{zhou2025mr}, whereas integrating visual modalities introduces inferential complexity. Recent multimodal benchmarks~\cite{zhou2025mr, zhang2025mrmr} have extended RIR to vision-language scenarios, revealing a pronounced gap in current MLLMs~\cite{jiang2025ev, gme2025zhang}, when tasked with reasoning over joint visual-textual evidence (\eg spatial relations, causal structure). 
From a retriever-capability perspective, progress depends on perceptually faithful and fine-grained phrase-to-region grounding, compositional representations that encode explicit cross-model rationales, 
and the ability to aggregate reasoning across  interleaved multi-image evidence.

\paragraph{Inference Latency and Cost.}
Many high-performing approaches rely on complex multi-stage~\cite{long2025diver} or reasoning-enhanced~\cite{chen2025reasonembed} pipelines, resulting in high inference costs. 
This issue partly stems from the limited reasoning capacity of compact embedding representations and the constraints introduced by contrastive learning.
Developing methods (\eg latent reasoning~\cite{jin2026laserinternalizingexplicitreasoning} or multi-vector representations~\cite{colbert2020omar}) that balance effectiveness and efficiency would significantly improve practical deployment. More broadly, adaptive routing can allocate reasoning budget based on query difficulty or scenario to control cost without uniformly sacrificing quality.

\paragraph{Generalization Bottlenecks and Narrow Application Scope.}
Although several works demonstrate cross-benchmark generalization by evaluating on both RIR and traditional IR benchmarks, these specialized methods still often underperform compared to strong general-purpose embedding models~\cite{geminiemb, zhang2025qwen3embeddingadvancingtext, jinaemb}. 
Furthermore, current RIR research mainly focuses on retrieval and reranking. However, RIR naturally aligns with broader applications such as long-term memory systems and deep research assistants. For instance, when a scientist asks a complex research question, a reasoning retriever could leverage historical interests and prior publications to provide personalized and context-aware evidence. Expanding RIR to these practical scenarios presents both evaluation and methodological opportunities.

\section{Conclusion}
This survey provides a structured roadmap for the rapidly evolving field of RIR. It systematizes the fragmented landscape of benchmarks and datasets, providing a detailed characterization of their difficulty, knowledge domains, and modalities, 
while introducing a comprehensive reasoning-type taxonomy with examples and an analysis of the reasoning-type focus for each benchmark. 
We introduce a fine-grained taxonomy that organizes approaches based on where reasoning is incorporated into the retrieval pipeline, 
spanning pre-retrieval augmentations, retriever training, advanced reranking, and iterative workflows.
To contextualize these paradigms, we synthesize theoretical analyses of optimization objectives and provide empirical comparisons for performance and model backbone, mapping these methods to relevant tasks and applications.
Finally, we identify key challenges including evaluation metrics innovation, generalization bottlenecks in evaluation and methodologies, bridging the multimodal reasoning gap, and alleviating inference computational costs to make LLM-driven reasoning practical. Addressing these issues is essential for developing the next generation of search systems that are generalizable, reasoning-capable, and practically deployable at scale.

\section*{Limitations}
While this survey provides an up-to-date and comprehensive review on reasoning-intensive retrieval, we acknowledge several limitations of this survey. 
First, we mainly include methods that have been empirically evaluated on established reasoning-intensive retrieval benchmarks. Other promising directions (\eg graph-based retrieval and Hypothetical Document Embeddings, HyDE) are not discussed in depth. 
Second, our review is restricted to publicly accessible literature and resources, which may overlook proprietary systems and unpublished industrial advances.


\appendix

\clearpage
\begin{table*}[h]
\centering
\small
\resizebox{\textwidth}{!}{
\begin{tabular}{p{1.4cm} L{2.6cm} r L{2.2cm} L{2.2cm} L{3cm} L{5cm}}
\toprule
\textbf{Domain} & \textbf{Name} & \textbf{Size} & \textbf{Data Source} & \textbf{Query} & \textbf{Doc} & \textbf{Reflecting Real-World Difficulty} \\
\midrule
\multirow{2}{*}{\parbox{1.4cm}{\centering\textbf{Open Domain}}}
& \benchname{ImpliRet}{taghavi-etal-2025-impliret} & 9,000 & Internet, LLM & Natural Text & Chat history & Document-side reasoning with no lexical overlap \\
& \benchname{BESPOKE}{kim2025bespoke} & 150 & Human & Natural Text & Chat history & Capture implicit user preferences in multi-turn chat APP \\
\midrule
\multirow{4}{*}{\parbox{1.4cm}{\centering\textbf{Scientific}}}
& \benchname{MIRB}{ju2025mirb} & 39,029 & Internet, Math Libraries, Previous Dataset & Natural/Formal Text & Theorem, Formula, Proof, Question & Automated math theorem proving \\
& \benchname{MathNet-Retrieve}{alshammari2025mathnet} & 10,000 & Contest, Human, LLM & Formal Text/Image & Similar Question  & Retrieve mathematically equivalent problems in multilingual and multimodal domains.\\
& \benchname{ScIRGen}{lin2025scirgen} & 61,376 & Internet, LLM, Papers & Natural Text & Paper Content & Complex task-oriented research questions in scientific workflows \\
& \benchname{FreshStack}{thakur2025freshstack} & 672 & Internet, LLM & Natural Text & Document, Code & Find realistic solutions from niche, up-to-date technical documents \\
\midrule
\multirow{2}{*}{\parbox{1.4cm}{\centering\textbf{Code}}}
& \benchname{CoIR}{li-etal-2025-coir} & $\approx$162,000 & Contest, Human, Internet, LLM, Previous Dataset & Natural Text, Code & Code, Answer & Code Summary, Code Translation \\
& \benchname{CoQuIR}{geng2025coquir} & 42,725 & Internet, LLM, Previous Dataset & Natural text & Code & Prioritizing quality over mere functional relevance \\
\midrule
\multirow{1}{*}{\parbox{1.4cm}{\centering\textbf{Legal}}}
& \benchname{Legal-Benchmark}{legalbenchmark2025} & 9,863 & Databases, Human, Internet, Textbooks & Natural Text & Answer, Statute & Quick search for relevant statutes based on realistic legal issues. \\
\midrule
\multirow{2}{*}{\parbox{1.4cm}{\centering\textbf{Medical}}}
& \benchname{R2MED}{li2025r2med} & 876 & Human, Internet, LLM, Papers, Previous Dataset, Textbooks & Natural Text & Answer, Document, Diagnosis & Explore complete latent diagnoses and treatment planning from symptoms for doctors \\
& \benchname{CMIRB}{li-etal-2025-automir} & 10,962 & Internet, LLM, Papers, Previous Dataset & Natural Text & Document, Diagnosis, Question & Match patient symptoms to consultations \\
\midrule
\multirow{3}{*}{\parbox{1.4cm}{\centering
\textbf{Multi-Domain}\\
}}
& \benchname{BRIGHT}{su2025bright} & 1,384 & Contest, LLM, Human, Internet, Previous Dataset, Textbooks & Natural Text & Theorem, Code, Document, Question & Find supportive evidence with deeper logical connection (\eg scientific search) \\
& \benchname{BRIGHT+}{chen2025brightplus} & 1,384 & Contest, LLM, Human, Internet, Previous Dataset, Textbooks & Natural Text & Theorem, Code, Document, Question & (same as BRIGHT) \\
& \benchname{RAR-b}{xiao2024rarb} & 45,745 & Internet, Previous Dataset & Natural Text & Answer, Code & Automated answer annotation on scientific QA \\
\midrule
\multirow{4}{*}{\parbox{1.4cm}{\centering\textbf{Multi-Modal}}}
& \benchname{MRMR}{zhang2025mrmr} & 1,435 & LLM, Human, Internet, Previous Dataset & Natural Text/Image & Answer, Theorem, Document, Image & Expert-level visual interpretation and interleaved modalities \\
& \benchname{MR2-BENCH}{zhou2025mr} & 1,309 & LLM, Human, Internet, Papers, Previous Dataset, Textbooks & Natural Text/Image & Document, Theorem, Diagram, Image & Understand and retrieve content in complex and multi-modal document structure. \\
& \benchname{ARK}{lin2026arkdualaxismultimodalretrieval} & 1,547 & LLM, Human, Internet, Previous Dataset, Papers & Natural Text/Image & Image, Diagram, Chart, Scientific Illustration & Abstract conceptual connection between knowledge and scientific documents. \\
& \benchname{MM-BRIGHT}{abdallah2026mmbrightmultitaskmultimodalbenchmark} & 2,803 & LLM, Human, Internet & Natural Text/Image & Document, Image, Multimodal & Reasoning-intensive multi-task retrieval from real expert technical queries with integral images. \\
\bottomrule
\end{tabular}
}
\caption{Overview of existing reasoning-intensive retrieval benchmarks. 
}
\label{tab:benchmarks}
\end{table*}

\section{Literature Review Procedure}
\label{app:reviewprocedure}
To ensure transparency and rigor, we provide the paper collection strategy and paper selection strategy in this section. 

\paragraph{Databases. } We searched major sources including ACL Anthology, OpenReview (ICLR, NeurIPS), arXiv, Semantic Scholar, DBLP, and Google Scholar, Github.  AI search tools such as PASA, Litmap Connected Papers are also included. 

\paragraph{Search Strategy. } We applied keyword combinations such as ``reasoning retriever,'' and ``retrieval reasoning,'' within the time range 2024--2026. In addition, we adopt a snowballing strategy by tracing the references and citations of seminal works (\eg BRIGHT~\cite{su2025bright}) and recent contributions (\eg R2Med~\cite{li2025r2med}, ReasonIR~\cite{shao2025reasonir}). 

\paragraph{Paper Selection Strategy.  }  
We have two main principles for selecting suitable papers:
(1) The paper must directly address both reasoning and retrieval or search.
(2) The paper must be publicly available as a journal article, conference paper, or preprint.
Additionally, we will not select papers that:
(1) Are abstracts, short articles, or non-academic blog posts.
(2) Do not have an accessible full text. 

\paragraph{Screening and Statistics.} 
Our initial screening retrieved approximately 400 articles. After deduplication, around 300 articles remained. 
Applying the inclusion criteria and exclusion yielded 118 papers.
After careful human validation of each paper, we finally selected out 56 qualified papers in this domain.  
\paragraph{Methodological Rigor.} 
Our protocol is informed by established guidelines for systematic reviews.  
These emphasize transparent reporting of search strings, last search date, de-duplication, per-stage counts, and inclusion/exclusion flows. By following these standards, we ensure that our literature review process is rigorous, reproducible, and aligned with recognized best practices.

\section{Reasoning Type Definition}
\label{app:reasoning_types}

The following reasoning paradigms characterize how domain knowledge in documents supports query resolution. Each type is defined with its logical mechanism and concrete examples. 

\paragraph{Deductive Reasoning. } A general principle or theorem in the document is directly applied to explain a specific scenario or solve a problem in the query. \textit{Example:} In ~\autoref{tab:domain-benchmarks} (General/BRIGHT), meristem regeneration theory explains post-cut tree sprouting.

\paragraph{Analogical Reasoning. } A document draws a parallel with the query in its underlying logic, indicating that the query and document share a solution strategy or a common theorem/algorithmic foundation.
\textit{Example:} In Table~\ref{tab:domain-benchmarks} (Code/COIR), a C++ Levenshtein distance implementation guides a Python solution via algorithmic equivalence. 

\paragraph{Causal Reasoning. } The document identifies root causes or mechanistic relationships that explain effects observed in the query. Resolution requires tracing causal chains from symptoms to origins. \textit{Example:} In Table~\ref{tab:domain-benchmarks} (Code/BRIGHT), missing debug messages are traced to launch file log-level configurations.

\paragraph{Analytical Reasoning. } The document provides critical domain knowledge that fills gaps in multi-step reasoning chains required to resolve the query. This involves decomposition of complex problems into interdependent sub-questions. \textit{Example:} In Table~\ref{tab:domain-benchmarks} (General/BRIGHT), soil science knowledge about salt accumulation completes the reasoning chain for plant water reuse safety. 

\paragraph{Numerical Reasoning. } The query is resolved by applying quantitative constraints in the document, requiring arithmetic computation (\eg percentages, unit conversion, rate/ratio) or time arithmetic (\eg duration, scheduling offsets, temporal comparisons). The logical mechanism is a deterministic mapping from numeric facts and rules to a target value or decision. \textit{Example:} In Table~\ref{tab:domain-benchmarks} (General/ImpliRet), the document states ``Prada Galleria costs \$2{,}000'' and ``Gucci Marmont is 20\% cheaper,'' so computing $2000 \times 0.8 = 1600$ identifies the Gucci Marmont as the \$1{,}600 bag. 
\section{Complex Retrieval Tasks}
\label{app:complex_retrieval}

To rigorously define the scope of Reasoning-Intensive Retrieval, it is essential to distinguish it from other established complex retrieval paradigms. While these tasks share the need for capabilities beyond simple keyword matching, they differ fundamentally in their core objectives and the nature of the query-document connection.
\subsection{Types and Definitions}
\paragraph{Multi-Hop Retrieval}
Multi-hop retrieval addresses scenarios where answering a query requires finding a chain of supporting facts~\cite{yang-etal-2018-hotpotqa}. The questions are mostly complex that no single document can resolve (\eg Document A mentions an entity $X$, and Document B provides the target attribute of $X$).

\paragraph{Instruction-Following Retrieval}
Instruction-following retrieval evaluates a retriever's ability to adhere to complex, explicit constraints provided in the user query~\cite{oh2024instructirbenchmarkinstructionfollowing}. For example, detailed directives regarding length, format, style, or negative constraints (\eg retrieve documents about apples but exclude any mention of technology). 

\paragraph{Long-Context Retrieval}
Long-context retrieval focuses on the challenge of identifying relevant information (``needles'') buried within extremely long inputs (``haystacks''), such as entire books or long legal contracts~\cite{zhu-etal-2024-longembed}, aiming to test the fidelity of embedding models over extended sequence lengths (\eg 32k+ tokens) . The core difficulty lies in the \textit{scale} of the context rather than the complexity of the reasoning. 

\subsection{Comparison with RIR}
While complex retrieval tasks  involve intricate constraints, they often rely on features explicitly specified in the query (\eg specific entity attributes in multi-hop retrieval, formatting constraints in instruction-following retrieval). Consequently, these tasks can often be addressed through precise lexical matching (\eg BM25) or surface-level semantic alignment. In contrast, Reasoning-Intensive Retrieval is defined by relevance signals that are mediated through \textit{latent logical inference chains} (see examples in \autoref{tab:domain-benchmarks}). Because the connection is implicit rather than explicitly stated, RIR necessitates a retriever capable of performing reasoning to bridge the gap, rather than relying solely on surface-level overlap.

\section{Empirical Analysis of RIR Methods.}
\label{app:method_comparison}

Beyond categorizing RIR methods, evaluating their practical deployability requires analyzing their empirical performance. This section analyzes the inherent trade-offs between computational overhead, reasoning capacity, and downstream ranking effectiveness. Specifically, we compare the roles of different base models and examine the steep scaling costs associated with multi-stage inference. 

\paragraph{LLM-Based vs. LRM-Based Methods.}

Large Reasoning Models (LRMs) are more suitable for ``thinking-heavy'' stages, such as complex query rewriting \cite{deepseek2025} and reranking \cite{liu2025reasonrank}, where deeper multi-step inference is required and slightly higher latency is tolerable. In contrast, standard LLMs typically serve as the backbone of the core retrieval stage due to stricter latency constraints, where efficiency and scalability are critical. However, LRMs remain critical offline; they curate high-quality, reasoning-intensive training data to fine-tune standard retrievers to better capture latent logical relevance \cite{long2025diver}. Furthermore, some frontier approaches \cite{zhang2025youradaqr, jin2026laserinternalizingexplicitreasoning} have also explored transferring reasoning capabilities from LRMs to LLM architectures through techniques such as distillation, aiming to achieve a better trade-off between effectiveness and efficiency.

\paragraph{Computation Cost vs. Performance.}
~\autoref{tab:bright_comparative} and ~\autoref{tab:r2med_comparative} summarize the framework and performance among methods. 
Additionally, we provide computational overhead across different methods in ~\autoref{tab:bright_comparative}. 

To compare the efficiency of different models, we follow the closed-form formulation of E2R-FLOPs~\cite{er2metric} and instantiate the cost using each model's architectural hyperparameters, including the number of layers, hidden size, feed-forward dimension, and attention configuration. For single-vector embedding backbones, we estimate the cost of one forward pass using an effective input length defined as the average of the query length and document length, i.e., $(L_q + L_d)/2$, which reflects the mean encoding cost under our corpus statistics. 
For reranking backbones, we estimate prompt-side FLOPs according to the reranking paradigm: pointwise methods process one query-document pair per call, groupwise and setwise methods process groups of five documents per call, and listwise methods process windows of twenty documents per call. The total computational overhead is then obtained by multiplying the per-call FLOPs by the corresponding number of calls required to rank the top candidates. In this way, our comparison normalizes efficiency across heterogeneous backbones and inference strategies under a unified, hardware-agnostic FLOPs metric. 

Foundational single-stage dense retrievers, operating via standard dot-product scoring, deliver robust performance in both effectiveness and robustness. Notably, the strongest retriever, ReasonEmbed-8B~\cite{chen2025reasonembed} even outperforms a $4\times$ larger reranker ReasonRank-32B~\cite{liu2025reasonrank} on both benchmarks. 

However, maximizing effectiveness often requires transitioning to multi-stage reasoning architectures. 
For instance, adding reasoning-aware rerankers yields higher nDCG scores, and multi-step agentic pipelines even achieve  peak metrics. On the other hand, they escalate inference costs to the $10^{14}$ and $10^{16}$ FLOPs respectively. 
Thus, frontier approaches seek a middle ground to enhance performance while bounding or reducing compute. In the reranking domain, GroupRank~\cite{sun2025grouprank}  combines pointwise computational efficiency with listwise contextual effectiveness. Furthermore, within multi-stage pipelines, INF-X-Retriever~\cite{inf-x-retriever-2025} achieves state-of-the-art performance without compute-heavy rerankers, by directly pairing an intent-recognizing query aligner with a highly optimized retriever. 
\section{Loss Function}
\label{app:loss_function}
\paragraph{InfoNCE (Information Noise-Contrastive Estimation).}
InfoNCE is a standard objective for self-supervised contrastive learning. Given a query embedding $q$, a matched positive document $d^{+}$, and a set of negatives $D^{-}$ (optionally including hard negatives), the loss is
\begin{equation}
\label{eq:infonce}
\mathcal{L}_{\mathrm{InfoNCE}}
=
-\log
\frac{\exp\!\big(s(q,d^{+})/\tau\big)}
{\sum\limits_{d\in \{d^{+}\}\cup D^{-}} \exp\!\big(s(q,d)/\tau\big)} \, .
\end{equation}
where $s(\cdot,\cdot)$ denotes a similarity score and $\tau>0$ is a temperature hyperparameter.

InfoNCE trains retrievers to minimize the representation distance between relevant pairs (\eg logically related documents in RIR). Curating a high-quality, reasoning-intensive dataset is therefore essential for effective optimization. Specifically, hard negatives are critical for teaching the model to penalize documents that possess surface-level semantic relevance but are logically unrelated to the query~\cite{long2025diver, shao2025reasonir}. 

\paragraph{Generation Loss.}
In multi-task training for LLM-based retrievers, a generation objective is commonly used to produce intermediate thoughts or reasoning traces conditioned on the query. For instance $i$, let $q_i$ be the query/prompt tokens and $t_i$ the target thought tokens; define $x_i=[q_i;\,t_i]$ with $L_i=|q_i|+|t_i|$. The loss typically supervises only the target span:
\begin{equation}
\label{eq:gen_loss}
\mathcal{L}_{\mathrm{gen}}
=
-\sum_{i=1}^{N}\ \sum_{j=|q_i|+1}^{L_i}
\log p_{\theta}\!\left(x_{i,j}\mid x_{i,<j}\right),
\end{equation}
where $x_{i,<j}=(x_{i,1},\ldots,x_{i,j-1})$.

While standard generation loss optimizes autoregressive next-token prediction, recent LLM-based retrievers repurpose this objective to explicitly train intermediate reasoning steps~\cite{tang2025largelrem, lan2025ume}. Furthermore, LaSER~\cite{jin2026laserinternalizingexplicitreasoning} advances this approach by internalizing these reasoning patterns directly into the latent embedding space.
\paragraph{Mean Squared Error (MSE).}
MSE is commonly used for representation matching (\eg embedding distillation). Given input embeddings $e_i\in\mathbb{R}^d$ and target embeddings $e_i^{\star}\in\mathbb{R}^d$, a parametric mapping $\mathcal{M}(\cdot;\theta)$ is trained by
\begin{equation}
\label{eq:mse}
\mathcal{L}_{\mathrm{MSE}}
=
\frac{1}{M}\sum_{i=1}^{M}\left\lVert \mathcal{M}(e_i;\theta)-e_i^{\star}\right\rVert_2^2 .
\end{equation}

MSE helps distilling an LLM’s deep reasoning capabilities into a computationally cheap embedding space. By training a compact mapper to minimize the distance between a raw query's embedding and its LLM-reasoned counterpart, the system internalizes the semantic transformations of multi-step inference~\cite{zhang2025youradaqr}.

\section{Relevant Tasks and Applications}
\label{app:application_task}
Reasoning-intensive retrieval extends IR from superficial lexical / semantic relevance to latent inferential link, providing logically grounded evidence for complex tasks.
For example, in users' intent recognition task,  it improves aligning  implicit user query and target corpus by detailed reasoning thought~\cite{chen2026agentirreasoningawareretrievaldeep, zhu-etal-2025-convsearch}. Additionally, as a part of RAG, it enhances RAG performance by retrieving high-quality documents for truth grounding~\cite{shao2025reasonir}. In knowledge-intensive domains, it improves misinformation detection~\cite{yu2025truth}, fact check~\cite{liu2025suceareasoningintensiveretrievaladversarial}, scientific paper research~\cite{garikaparthi-etal-2025-mir},  complex QA~\cite{liu2025reasonrank}, contextual relevance judgment~\cite{ji2025reasontorank, huang2025contextualrelevanceadaptivesampling}, grounding responses in retrieved
knowledge and thereby mitigating hallucinations.

\noindent
Reasoning-intensive IR is increasingly applied across
diverse domains, including healthcare, software engineering, and e-commerce. The following sections explore domain-specific adaptations of these
techniques in greater depth.

\paragraph{Medicine}
Addressing the complexities of reasoning-intensive retrieval in the medical domain, the 
RAR$^{2}$~\cite{xu-etal-2025-rar2} framework improves diagnostic accuracy by generating an intermediate ``thought process'' that uncovers implicit clinical knowledge requirements to explicitly guide both the retrieval of evidence and the subsequent reasoning generation. 

\paragraph{E-Commerce}
In the realm of e-commerce reasoning-intensive retrieval,
LREM~\cite{tang2025largelrem} leverages reasoning-then-embedding approach effectively links implicit user queries with intended products, leading to more precise and meaningful retrieval. Additionally, LREF~\cite{lrefjingdong} optimizes retrieval performance by utilizing reasoning processes to achieve a more meticulous and granular alignment of query-product relevance. 

\paragraph{Software Engineering}
To address the intricacies of software engineering, reasoning-intensive retrieval improves performance by shifting from static semantic matching to a dynamic process of structural code exploration and verified algorithmic reasoning.
CR-Planner~\cite{li-etal-2025-elicit} significantly improves performance on rigorous tasks like competitive programming by employing a critic-guided planning framework to iteratively validate and refine both retrieval queries and reasoning steps, ensuring that generated code is grounded in accurate, verified evidence
LATTICE~\cite{gupta2025llmlattice} addresses the scalability challenges of searching massive software repositories by imposing a semantic tree structure on the corpus, enabling the LLM to actively traverse hierarchical paths and efficiently pinpoint deeply nested logic that flat retrieval methods often miss.

\begin{table*}[htbp]
\centering
\footnotesize 
\renewcommand{\arraystretch}{1.4} 
\setlength{\tabcolsep}{3pt} 

\begin{tabularx}{\textwidth}{l p{1.8cm} p{1.8cm} X X}
\toprule
\textbf{Domain} & \textbf{Benchmark} & \textbf{Reasoning Type} & \textbf{Example} & \textbf{Inference Chain} \\
\midrule

\multirow{3}{*}{\textbf{Open Domain}} 
& ImpliRet
& Numerical
& \textbf{Query:} ``Which bag costs \$1,600?'' \newline
  \textbf{Related Doc:} ``The Prada Galleria costs \$2,000; the Gucci Marmont is 20\% cheaper.''
& Given reference price (\$2,000) \newline
  $\to$ Apply discount rule (20\% cheaper $\Rightarrow 0.8\times 2000= 1600$) \newline
  $\to$ Match target amount (\$1,600 $\Rightarrow$ Gucci) \\
\midrule

\multirow{9}{*}{\textbf{Scientific}} & MIRB & Deductive \newline (Symbolic) & \textbf{Query:} ``Open covering H of closed bounded S in R has finite subcover He from H'' \newline \textbf{Related Doc:} No point in S$^c$ is limit point of S & Theorem (Heine-Borel for compactness) \newline $\to$ Prerequisite (S closed and bounded) \newline $\to$ Property (closed: no exterior limit points) \\
\cmidrule{2-5}
& MathNet-Retrieve & Analytical \newline (Multimodal) & \textbf{Query:} ``Prove points D, E, F, G, H are concyclic...'' \newline \textbf{Related Doc:} Proof by drawing EG and FH, chasing equal angles... & Core problem $\to$ Analysis root (parallelogram) $\to$ Implications (parallel lines) $\to$ Conclusion (equal angles force concyclicity) \\
\cmidrule{2-5}
& BRIGHT \newline (Biology) & Deductive & \textbf{Query:} ``After cutting trees into logs... they grow normal stems...'' \newline \textbf{Related Doc:} Document on meristematic tissues & Phenomenon $\to$ Supportive theory (cell division) $\to$ Applied concept (meristem) \\
\midrule

\multirow{5}{*}{\textbf{Code}} & BRIGHT \newline (Robotics) & Causal & \textbf{Query:} ``Can’t see debug messages using RCLCPP\_DEBUG...'' \newline \textbf{Related Doc:} Launch file with log\_level default 'info'... & Symptom $\to$ Potential cause (node log level override) $\to$ Configuration (default 'info' arg) \\
\cmidrule{2-5}
& COIR & Analogical & \textbf{Query:} Python code implementing Levenshtein distance... & Query pattern $\to$ Algorithmic equivalence $\to$ Language translation $\to$ Structural mapping \\
\midrule

\textbf{Legal} & LegalBench & Deductive & \textbf{Query:} ``Teacher fired from private school...'' \newline \textbf{Related Doc:} 14th Amendment Due Process... & Legal issue $\to$ Supportive Rule $\to$ Rules application $\to$ Facts connection $\to$ Conclusion \\
\midrule

\multirow{4}{*}{\textbf{Medical}} & R2MED & Analytical & \textbf{Query:} ``An 82-year-old woman... What is the next test?'' \newline \textbf{Related Doc:} Video-capsule endoscopy... & Core problem $\to$ Analysis root $\to$ Latent reasoning $\to$ Diagnostic method \\
\cmidrule{2-5}
& CMIRB & Deductive & \textbf{Query:} ``How long after thyroid surgery can one return to work?'' \newline \textbf{Related Doc:} Healing timeline... & Phenomenon $\to$ Supportive healing process $\to$ Relevant timeline \\
\midrule

\multirow{6.5}{*}{\textbf{Multimodal}} 
& MRMR 
& Deductive 
& \textbf{Query:} ``Jack was driving through...'' \newline
  \textbf{Image:} A white car crossing lane... \newline
  \textbf{Related Doc:} ``Driving in tunnels — Rule (f)''
& Observed behavior $\to$ Applicable regulation $\to$ Constraint violation $\to$ Relevant document retrieval \\
\cmidrule{2-5}
& MRMR & Causal & \textbf{Query:} ``What causes black bulges on a corn cob?'' & Visual symptom $\to$ Potential cause $\to$ Specific cause $\to$ Disease identification \\
\bottomrule
\end{tabularx}
\caption{Examples of domain-specific benchmarks with key reasoning types, query examples, and inference chains.}
\label{tab:domain-benchmarks}
\end{table*}

\begin{table*}[t]
    \centering
    \small
    \renewcommand{\arraystretch}{1.35}   
    \setlength{\tabcolsep}{4pt}
    \begin{tabular}{l @{\hspace{2pt}} >{\raggedright\arraybackslash}p{2.5cm} >{\raggedright\arraybackslash}p{1.9cm} c 
                    >{\raggedright\arraybackslash}p{3.0cm} l c r}
    \toprule
    \textbf{Role} & \textbf{Method} & \textbf{Backbone} & \textbf{Size} & \textbf{Framework} & \textbf{Inference} & \textbf{nDCG@10} & \textbf{FLOPs} \\
    \midrule

    \multicolumn{8}{l}{\textit{\textbf{Group 1: Single-Stage Retrieval}}} \\
    \midrule
    \rowcolor{rowblue}
    & & & 8B & Single Vector & Dot Product & \textbf{38.1} & 2.0584e12 \\
    \rowcolor{rowblue}
    & \multirow{-2}{2.5cm}{ReasonEmbed \footnotesize\cite{chen2025reasonembed}} 
      & \multirow{-2}{1.9cm}{Qwen-3} & 4B & Single Vector & Dot Product & 37.1 & 1.1001e12 \\
    & DIVER-Retriever \footnotesize\cite{long2025diver} 
      & Qwen-3 & 4B & Single Vector & Dot Product & 28.9 & 1.1001e12 \\
    \rowcolor{rowblue}
    & RaDeR \footnotesize\cite{das-etal-2025-rader} 
      & GTE-Qwen2 & 7B & Single Vector & Dot Product & 25.5 & 1.8511e12 \\
    \multirow{-5}{*}{\textbf{Retriever}} 
      & ReasonIR \footnotesize\cite{shao2025reasonir} 
      & Llama-3.1 & 8B & Single Vector & Dot Product & 24.4 & 2.0443e12 \\
    \midrule

    \multicolumn{8}{l}{\textit{\textbf{Group 2: Reranking \& Multi-Stage Pipelines}}} \\
    \midrule
    \rowcolor{rowblue}
    & & & 32B & DIVER-4B + GPT-4 query rewrite & CoT Generation & \textbf{39.2} & 2.2687e15 \\
    \rowcolor{rowblue}
    & \multirow{-2}{2.5cm}{GroupRank \footnotesize\cite{sun2025grouprank}} 
      & \multirow{-2}{1.9cm}{Qwen-2.5} & 7B & DIVER-4B + GPT-4 query rewrite & CoT Generation & 36.7 & 4.7405e14 \\
    & & & 32B & ReasonIR-8B + GPT-4 query rewrite & CoT Generation & 38.0 & 2.2207e15 \\
    & \multirow{-2}{2.5cm}{ReasonRank \footnotesize\cite{zhang-etal-2025-rearank}} 
      & \multirow{-2}{1.9cm}{Qwen-2.5} & 7B & ReasonIR-8B + GPT-4 query rewrite & CoT Generation & 35.7 & 4.6486e14 \\
    \rowcolor{rowblue}
    & Rank-R1 \footnotesize\cite{zhuang2025rank} 
      & Qwen-2.5 & 7B & with BM25 & CoT Generation & 16.4 & 4.7405e14 \\
    & & & 1.7B & with BM25 & Think-Free & 16.7 & 1.4556e14 \\
    \multirow{-7}{*}{\textbf{Reranker}} 
      & \multirow{-2}{2.5cm}{TFRank \footnotesize\cite{fan2025tfrank}} 
      & \multirow{-2}{1.9cm}{Qwen-3} & 0.6B & with BM25 & Think-Free & 15.6 & 5.1896e13 \\
    \cmidrule{2-8}

    \rowcolor{rowblue}
    & INF-X-Retriever \footnotesize\cite{inf-x-retriever-2025} 
      & GTE-Qwen2 & 7B & Query Aligner + Retriever & Multi-Step & \textbf{63.4} & - \\
    \multirow{-2}{*}{\textbf{Pipeline}} 
      & DIVER \footnotesize\cite{long2025diver} 
      & Qwen-3 & 8B & Expander + Retriever + Reranker & Multi-Step & 46.8 & - \\
    \bottomrule
    \end{tabular}
    \caption{Representative reasoning-intensive retrieval methods and their performance on the \textbf{BRIGHT} benchmark. Best score in each subgroup is in \textbf{bold}.}
    \label{tab:bright_comparative}
\end{table*}

\begin{table*}[t]
    \centering
    \begin{threeparttable}
    \small
    \renewcommand{\arraystretch}{1.3}
    \setlength{\tabcolsep}{8pt}
    \begin{tabular}{l >{\raggedright\arraybackslash}p{2.6cm} >{\raggedright\arraybackslash}p{2.2cm} c >{\raggedright\arraybackslash}p{4.0cm} r}
    \toprule
    \textbf{Role} & \textbf{Key Method} & \textbf{Backbone} & \textbf{Size} & \textbf{Framework Design} & \textbf{nDCG@10} \\
    \midrule
    
    \multicolumn{6}{l}{\textit{\textbf{Group 1: Retrieval} (Single-Stage Dense Retrieval)}} \\
    \midrule
    
    \rowcolor{rowblue}
    & & & 8B & Single Vector & \textbf{43.18} \\
    
    \rowcolor{rowblue} 
    & \multirow{-2}{2.6cm}{ReasonEmbed\newline {\footnotesize \cite{chen2025reasonembed}}} & \multirow{-2}{2.2cm}{Qwen-3} & 4B & Single Vector & 41.16 \\
    
    \multirow{-3}{*}{\textbf{Retriever}} & DIVER-Retriever\tnote{*}\newline {\footnotesize \cite{long2025diver}} & Qwen-3 & 4B & Single Vector & 42.91 \\
    \midrule
    
    \multicolumn{6}{l}{\textit{\textbf{Group 2: Reranking} (Multi-Stage)}} \\
    \midrule
    
    \rowcolor{rowblue}
    & & & 32B & with DIVER-Retriever-4B & \textbf{52.28} \\
    
    \rowcolor{rowblue} 
    & \multirow{-2}{2.6cm}{GroupRank\newline {\footnotesize \cite{sun2025grouprank}}} & \multirow{-2}{2.2cm}{Qwen-2.5} & 7B & with DIVER-Retriever-4B & 47.84 \\
    
    & & & 32B & with E5-mistral-7B & 42.85 \\
    
    \multirow{-4}{*}{\textbf{Reranker}} & \multirow{-2}{2.6cm}{ReasonRank\newline {\footnotesize \cite{zhang-etal-2025-rearank}}} & \multirow{-2}{2.2cm}{Qwen-2.5} & 7B & with E5-mistral-7B & 39.53 \\
    \bottomrule
    \end{tabular}
    
    \begin{tablenotes}
        \footnotesize
        \item[*] DIVER-Retriever data is from GroupRank paper.
    \end{tablenotes}
    \end{threeparttable}
    \caption{Representative Reasoning-Intensive Retrieval Methods Overview and Performance Landscape on \textbf{R2MED} benchmark. The top performance in each subgroup is highlighted in \textbf{bold}.}
    \label{tab:r2med_comparative}
\end{table*}

\end{document}